**Rotomagnetic couplings influence on the magnetic properties of antiferrodistortive antiferromagnets**

Eugene A. Eliseev[1], Maya D. Glinchuk[1], Venkatraman Gopalan[2] and Anna N. Morozovska[3]
[1] Institute for Problems of Materials Science, NAS of Ukraine,
Krjijanovskogo 3, 03142 Kyiv, Ukraine
[2] Department of Materials Science and Engineering, Pennsylvania State University, University Park, PA 16802, USA
[3] Institute of Physics, NAS of Ukraine, 46, pr. Nauki, 03028 Kyiv, Ukraine

Within the framework of Landau-Ginzburg-Devonshire (LGD) phenomenological theory we consider the possibility to control properties of paraelectric antiferromagnets via biquadratic rotomagnetic coupling with and without external magnetic and electric field application. We use $EuTiO_3$ as a prototype with relatively well-known material parameters. Surprisingly strong influence of this coupling practically on all the properties without external fields was obtained in the temperature region with coexistence of antiferromagnetic and antiferrodistorted phases i.e. in multiferroic state. In particular, the observed Neel temperature $T_N$ (5.5 K) was shown to be defined by rotomagnetic coupling, while without this coupling $T_N$ appeared to be much higher (26 K). For small or high enough rotomagnetic coupling constant value the antiferromagnetic phase transition order appeared to be the second or the first order respectively. The essential influence of rotomagnetic coupling on the form and value of magnetic and dielectric permittivity was also forecasted. The rotomagnetic coupling along with rotoelectric one opens the additional way to control the form of the phase diagrams by application of external magnetic or electric field. The critical value of the electric field (for antiferromagnetic to ferromagnetic phase transition) appeared essentially smaller than the one calculated without rotomagnetic coupling that can be important for applications.

## I. Introduction

### I.1. Definition of the roto-effects

Antiferrodistorted (AFD) perovskite oxides can possess octahedra oxygen rotations characterized by spontaneous octahedra tilt angles, which in turn can be described by an axial vector $\Phi_i$ (see a typical schematics in the **Figure 1).** Following Gopalan and Litvin [1] the AFD symmetry is in fact a "**rotosymmetry**" that includes 69 roto-groups. Typical AFD perovskites with octahedrally tilted phases are incipient ferroelectrics $SrTiO_3$, $CaTiO_3$, antiferromagnet incipient ferroelectric $EuTiO_3$, antiferromagnetic ferroelectric $BiFeO_3$, ferroelectric $Pb(Zr,Ti)O_3$ and antiferroelectric $ZrTiO_3$.

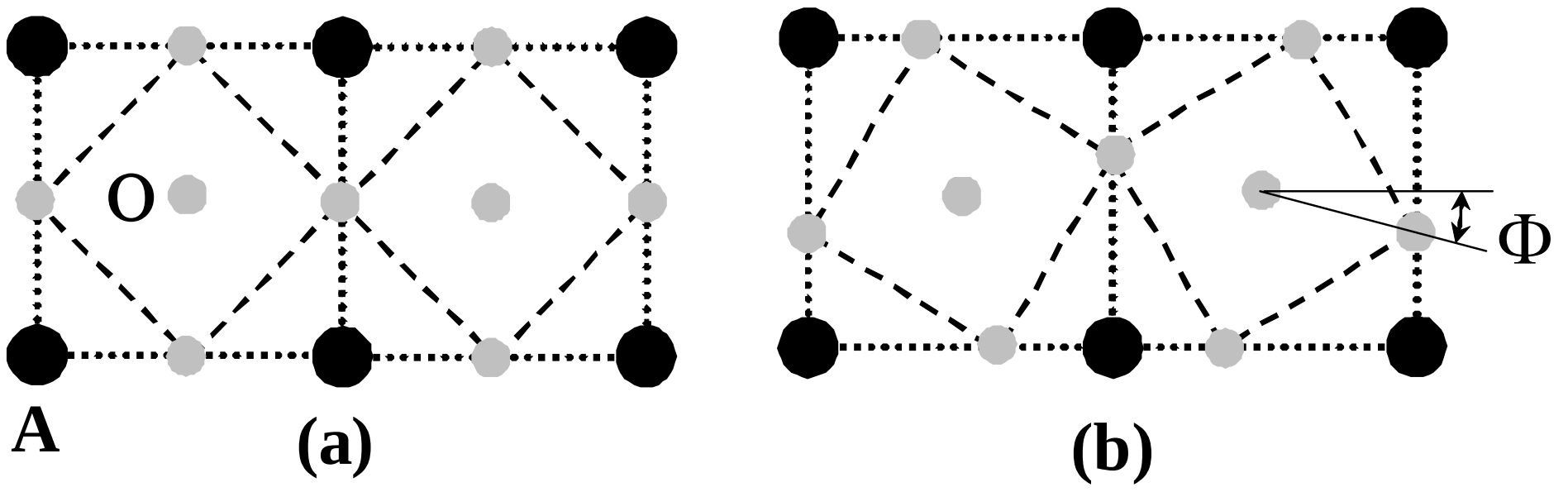

**Figure 1. (a)** Atomic ordering in $ABO_3$ perovskite structure in the nonstructural parent phase. **(b)** Antiferrodistortive (AFD) ordering in the structural phase is determined by the tilt $\Phi$. The tilt value is typically opposite for the neighbouring oxygen octahedrons $ABO_3$.

The structural, polar and magnetic orderings in AFD incipient ferroelectrics with cubic parent phase are linked directly via the different types of biquadratic coupling. For instance Balashova and Tagantsev [2] considered a multiferroic with two scalar order parameters coupled biquadratically and reported about its versatile phase diagrams. In particular, a spontaneous polarization vector $P_i$ can appear inside structural walls of $SrTiO_3$ and $CaTiO_3$ due to "**roto-electric**" (**RE**) biquadratic coupling term, $\xi_{ijkl}P_i P_j \Phi_k\Phi_l$ [3, 4]. The RE coupling term was later regarded as Houchmandazeh-Laizerowicz-Salje (HLS) coupling [5]. Biquadratic magnetoelectric (**ME**) coupling, described by the term $\eta_{ijkl}P_i P_j M_k M_l$ , was considered as the reason of magnetization appearance inside the ferromagnetic domain wall in a non-ferromagnetic media [6]. Surprising effects in the materials, where the "direct" rotomagnetic (**RM**) coupling described by the terms $\left(\xi^{M}_{ijkl}M_i M_j + \xi^{L}_{ijkl}L_i L_j\right)\Phi_k\Phi_l$ is included (**M** is magnetization, **L** is antiferromagentic order parameter) are considered in this work for the first time.

While all aforementioned biquadratic couplings (ME, RE and RM) are universal for the AFD materials with parent cubic symmetry, ME and RE coupling influence on their properties are relatively well studied, the RM coupling impact is almost terra incognita. To the best of our knowledge only one experiment in $EuTiO_3$ (ETO) that revealed magnetic field impact on the tilt was performed. Namely, Bussmann-Holder et al [7, 8] reported about the influence of magnetic field on the AFD phase transition temperature and magnetic susceptibility of ETO and observed slight influence in the vicinity of its AFD phase transition.

That is why below we will consider ETO with relatively well-known parameters as prototype of the possible group of paraelectric antiferromagnets. In particular the bulk quantum paraelectric ETO is a low temperature antiferromagnet [9, 10] with Neel temperature about 5.5 K. It exhibits an antiferrodistortive (AFD) transition at about 285 K [11, 12, 13, 14, 15] and is paraelectric at all temperatures.

Since the theory of RM coupling impact on the multiferroic properties was absent to date, the goal of the work is to propose the comprehensive LGD-based theory and to find out corresponding physical mechanisms of the RM coupling influence on phase diagrams and properties of incipient ferroelectric perovskites with oxygen octahedra rotations. The original part of the paper is organized as follows. The LGD functional and material parameters are given in Sec. II. RM effect in the vicinity of the AFD phase transition is analyzed in Section III. The impact of RM effects on the phase diagrams and physical properties at lower temperatures is demonstrated in Section IV. Sections V and VI are devoted to the discussion and conclusions respectively.

## II. LGD functional and material parameters

LGD approach is based on the phase stability analysis of thermodynamic potential that is a series expansion on powers of the order parameters, namely polarization P, sum and difference of sublattices magnetizations, *m* and *l*, and oxygen octahedra rotation angle Φ. For the case of homogeneous material with cubic high temperature parent phase, the bulk Gibbs potential is [16, 17]:

$$g = g_{FE} + g_{AFD} + g_{AFM} + g_{Coupling} \tag{1a}$$

Ferroelectric (FE) contribution to the energy (1a) is:

$$g_{FE} = \frac{\alpha_P}{2}P^2 + \frac{\beta_P}{4}P^4 - E_i P_i \tag{1b}$$

Here $P_i$ is ferroelectric polarization component, $P^2 = P_1^2 + P_2^2 + P_3^2$, $E_i$ is external electric field component. For incipient ferroelectric expansion coefficient $\alpha_P$ depends on the absolute temperature $T$ in accordance with Barrett law, namely $\alpha_P(T) = \alpha_T^{(P)}\left(T_q^{(P)}/2\right)\left(\coth\left(T_q^{(P)}/2T\right) - \coth\left(T_q^{(P)}/2T_c^{(P)}\right)\right)$. Here $\alpha_T^{(P)}$ is constant, temperature $T_q^{(P)}$ is the so-called quantum vibration temperature related with polar soft modes, $T_c^{(P)}$ is the "effective" Curie temperature corresponding to the polar soft modes in bulk quantum paraelectrics. Coefficient $\beta_P$ is regarded as temperature independent.

Antiferrodistortive (AFD) energy is

$$g_{AFD} = \frac{\alpha_\Phi}{2}\Phi^2 + \frac{\beta_\Phi}{4}\Phi^4 \qquad (1c)$$

The structural AFD order parameter $\Phi_k$ is the static tilt of oxygen octahedra, $\Phi^2 = \Phi_1^2 + \Phi_2^2 + \Phi_3^2$. The tilt vector expansion coefficient $\alpha_\Phi$ depends on the absolute temperature $T$. Usually one could use Barrett-type approximation for this coefficient temperature dependence in the form $\alpha_\Phi(T) = \alpha_T^{(\Phi)}\left(T_q^{(\Phi)}/2\right)\left(\coth\left(T_q^{(\Phi)}/2T\right) - \coth\left(T_q^{(\Phi)}/2T_S\right)\right)$(see e.g. ref.[18]), however the available experimental data [11, 12, 19 ,] do not allow us to determine unambiguously $T_q^{(\Phi)}$ parameter, hence we'll use high temperature limit of this expression in the form $\alpha_\Phi(T) = \alpha_T^{(\Phi)}(T - T_S)$.

Magnetic energy including magnetic order-disorder in the paramagnetic phase is [[20]]:

$$g_{AFM} = \frac{\alpha_M^{(T)} M_0^2}{2}\left[(T - T_C)m^2 + (T - T_N)l^2 + T\left\{S(m_i + l_i) + S(m_i - l_i) - m^2 - l^2\right\}\right] - M_0 H_i m_i \qquad (1d)$$

$M_0$ is the absolute value of saturation magnetization. We introduced two dimensionless order parameters, namely ferromagnetic (FM), $m_i = (m_{ai} + m_{bi})/2$, and antiferromagnetic (AFM), $l_i = (m_{ai} - m_{bi})/2$ ones with $m_{ai}$ and $m_{bi}$ as the components of dimensionless magnetizations of two equivalent sub-lattices. Magnetization $m^2 = m_1^2 + m_2^2 + m_3^2$ is square the dimensionless ferromagnetic order parameter, and $l^2 = l_1^2 + l_2^2 + l_3^2$ is the square of the dimensionless AFM order parameter absolute value, correspondingly. $H_i$ are magnetic field components. Configurational entropy is $S(m) = \frac{1+m}{2}\log\left(\frac{1+m}{2}\right) + \frac{1-m}{2}\log\left(\frac{1-m}{2}\right)$. Parameter $\alpha_M^{(T)}$ is related to magnetic Curie-Weiss constant $C_{CW}$ as $\alpha_M^{(T)} = \mu_0/C_{CW}$. $T_C$ is the seeding ferromagnetic Curie temperature and $T_N$ is the seeding Neel temperature for bulk material without antiferrodistotive ordering. In two sub-lattices antiferromagnets the negative $T_C$ value can be determined experimentally from inverse magnetic susceptibility in paramagnetic phase of the material. Examples of their observation one can find in [20].

Biquadratic coupling energy consists of magnetoelectric (ME), rotoelectric (RE) and rotomagnetic (RM) coupling terms and has the form:

$$g_{Coupling} = \frac{M_0^2}{2}\left(\eta_{FM} m^2 + \eta_{AFM} l^2\right)P^2 + \frac{\xi_{RE}}{2}\Phi^2 P^2 + \frac{M_0^2}{2}\Phi^2\left(\xi_{RM}^M m^2 + \xi_{RM}^L l^2\right) \tag{1e}$$

Following Lee *et al.* [21] we assume that the coefficients of FM and AFM order parameters contributions to ME coupling are equal by absolute value and have opposite signs, $\eta_{AFM} = -\eta_{FM}$. The biquadratic RE coupling coefficient $\xi_{RE}$ and RM coupling coefficients, $\xi_{RM}^M$ and $\xi_{RM}^L$, are regarded as temperature-independent [3, 4, 5]. Hereinafter we also regard that $\xi_{RM}^M = -\xi_{RM}^L$ as a consequence of two magnetic sub-lattices equivalence.

The equations of state for the dimensionless tilt $\phi = \Phi/\Phi_0$, magnetization $m = M/M_0$, antiferromagnetic order parameter $l = L/M_0$ and polarization $P$ were obtained from the minimization of the free energy (1). Here $\Phi_0 = \sqrt{\alpha_T^{(\Phi)} T_S/\beta_\Phi}$ is the value of tilt at low temperatures for the case of zero rotomagnetic coupling. Dielectric, magnetic and magnetoelectric susceptibility tensors can be calculated from expressions $\chi_{ij}^E = \partial P_i/\partial E_j$, $\chi_{ij}^M = \partial M_i/\partial H_j$ and $\chi_{ij}^{ME} = \partial P_i/\partial H_j$ correspondingly. These properties are tilt-dependent due to the RE and RM coupling. Expansion coefficients of the LGD potential from Eq.(1) used in our numerical calculations are listed in the **Table 1.** Most of the coefficients correspond to the bulk ETO, but ME, RE and RM coupling coefficients correspond to the prototype material and describe ETO properties semi-quantitatively.

.

**Table 1**. Expansion coefficients of the LGD potential from Eq.(1) used in numerical calculations.

| Coefficient | SI units | Value |
|---|---|---|
| Coefficient before $P^2$, $\alpha_P(T) = \alpha_T^{(P)}\left(T_q^{(P)}/2\right)\left(\coth\left(T_q^{(P)}/2T\right) - \coth\left(T_q^{(P)}/2T_c^{(P)}\right)\right)$ | | |
| Inverse dielectric stiffness constant, $\alpha_T^{(P)}$ | $10^6$ m/(F K) | 1.95 |
| Effective Curie temperature $T_c^{(P)}$ | K | -133.5 |
| Characteristic temperature $T_q^{(P)}$ | K | 230 |
| Coefficient before $P^4$, $\beta_P$ | $10^9$ $m^5/(C^2F)$ | 1.724 |
| Coefficient before $m^2$, $\alpha_M(T) = \alpha_T^{(M)}(T - T_C)$ | | |
| $\alpha_T^{(M)}$ | Henri/(m·K) | $4.36\cdot10^{-7}$ |
| FM Curie temperature, $T_C$ | K | -17.0 |
| Saturation magnetization, $M_0$ | A/m | $1.09\times10^6$ |
| Coefficient before $l^2$, $\alpha_L(T) = \alpha_T^{(M)}(T - T_N)$ | | |
| Inverse Neel constant, $\alpha_T^{(L)}$ | Henri/(m·K) | $4.36 \cdot 10^{-7}$ |
| AFM Neel temperature, $T_N$ | K | 26.2 |
| Coefficient before $\Phi^2$, $\alpha_\Phi(T) = \alpha_T^{(\Phi)}(T - T_S)$ | | |
| Coefficient $\alpha_T^{(\Phi)}$ | J/(m$^5$ K) | $4.184\times10^{26}$ |
| AFD transition temperature $T_S$ | K | 285 |
| Coefficient before $\Phi^4$, $\beta_\Phi$ | J/m$^7$ | $2.981\times10^{50}$ |
| ME coupling coefficient $\eta_{AFM}$ | J m$^3$/(C$^2$ A$^2$) | $8\times10^{-5}$ |
| ME coupling coefficient $\eta_{FM}$ | J m$^3$/(C$^2$ A$^2$) | $-8\times10^{-5}$ |
| Rotoelectric HLS ΦP-coupling coefficient $\xi_{RE}$ | (F m)$^{-1}$ | $-2.225\times10^{29}$ |
| Rotomagnetic ΦM-coupling coefficient $\xi_{RM}^{M}$ | N/(m$^2$A$^2$) | $-2.3\times10^{-16}$ |
| Rotoantiferromagnetic ΦL-coupling $\xi_{RM}^{L}$ | N/(m$^2$A$^2$) | $+2.3\times10^{-16}$ |

## III. RM effect in the vicinity of the AFD phase transition

As in was mentioned in the introduction, Bussmann-Holder et al [7, 8] observed the influence of the magnetic field on the AFD phase transition in ETO. In accordance with both their experiments and our calculations, magnetic permittivity should have the change of slope for the second order transition at the point of AFD transition (see **Figure 2**). Symbols in the **Figure 2** represent experimental results taken from Ref. [8]; our fitting is shown by the solid lines. Using known temperature dependence of tilt Φ and comparing the temperature dependence of permittivity above and below AFD transition one can determine the magnetic parameters, namely magnetic Curie temperature $T_C$ without and with RM coupling, -17 K and 3.8 K respectively, Curie-Weiss constant $C_{CW}$=2.878 K, and the coupling coefficient $\xi_{RM}^{M} = -2.3\times10^{16}$ N/(m$^2$A$^2$) [22].

Calculation results given in the **Figure 3a** shows that the external magnetic field causes the shift of AFD phase transition temperature to the higher temperatures and slightly increases AFD order parameter due to the RM coupling with $\xi_{RM}^{M} < 0$. Despite the temperature shift is rather small (about 0.3 K at 9 Tesla), it is in a reasonable agreement with the similar trend observed experimentally [7]. **Figure 3b** illustrates the "facture" (rapid change of the slope) of the dielectric permittivity temperature dependence below the AFD transition temperature induced by RM coupling in applied magnetic field.

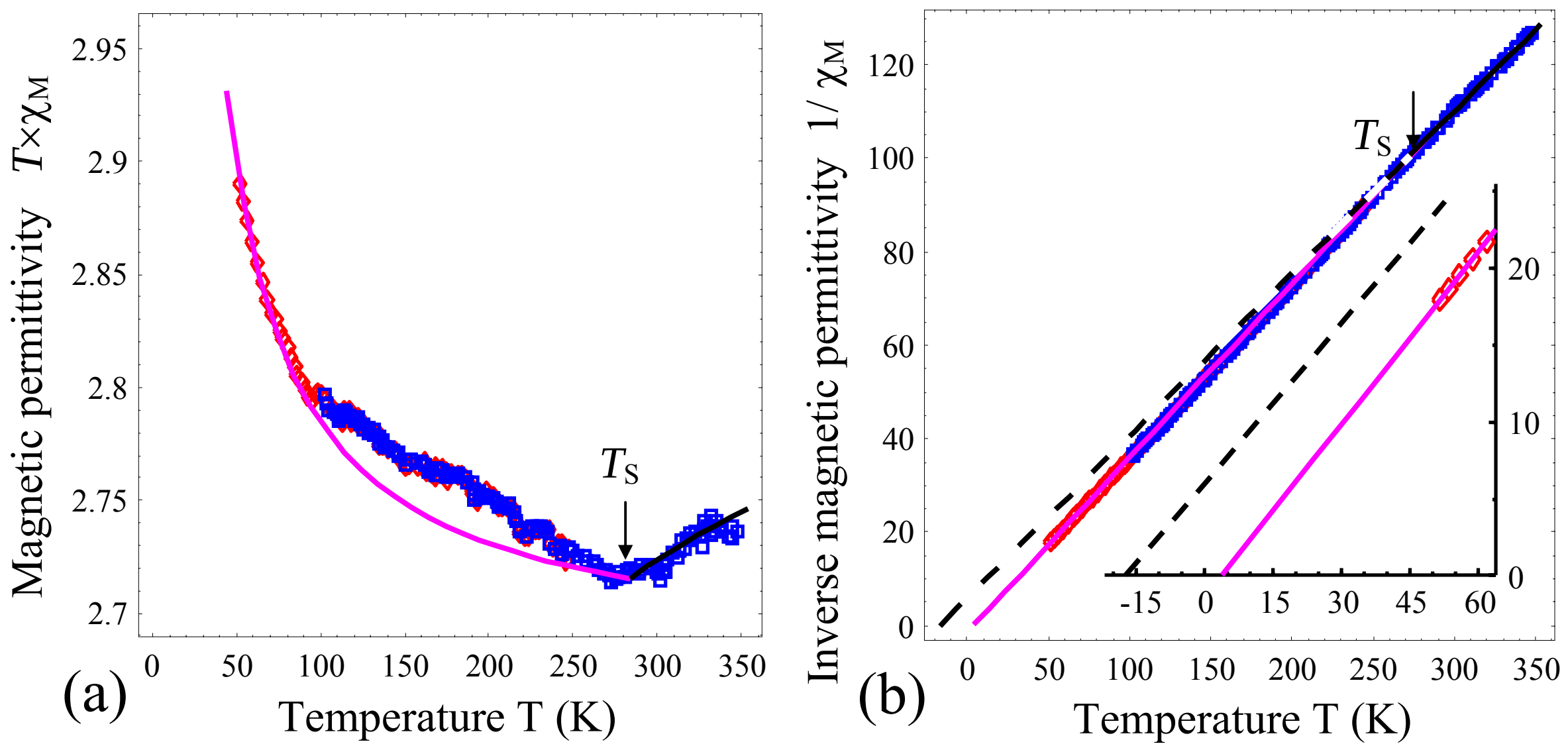

**Figure 2. Temperature dependence of ETO magnetic permittivity near the transition to AFD phase.** Symbols represent experimental results taken from Ref. [8] at very small measuring magnetic field *H*. Our fitting is for dimensionless magnetic permittivity multiplied by T **(a)** and inverse magnetic permittivity **(b)** are shown with solid curves for the following values of parameters: $T_S$= 285 K, $T_C$ = – 17 K, $C_{CW}$=2.878 K, $\xi^{M}_{RM} = -2.3\times 10^{16}\ \mathrm{N}/\left(\mathrm{m}^2\mathrm{A}^2\right)$. Dashed curve is the extrapolation of the dependence just above the transition temperature $T_S$ to lower temperatures. Inset represents the low temperature behavior of inverse permittivity.

As one can see from the figure, the dielectric permittivity becomes slightly magnetic-field dependent below the temperature of AFD phase transition (≈285 K), since AFD order parameter only slightly increases with magnetic field increase in the temperature region in accordance with the **Figure 3a**. Additional magnetic field dependence in all temperature region, including T>Ts, originates due to bi-quadratic ME coupling (see Eq.(1e)). So, it should be concluded, that the magnetic field dependence of the tilt indirectly increases the dependence of dielectric constant on the field due to the RE coupling between the tilt and polarization. ME coupling leads to the direct contribution to dielectric constant at all temperatures.

Effects shown in the **Figures 3** appeared in the immediate vicinity of AFD transition temperature (285 K) are small enough, since the AFD order parameter is small here. More pronounced and intriguing peculiarities of the dielectric susceptibility, magnetization and AFD order parameter can be caused by RM coupling at lower temperatures. Corresponding examples are demonstrated in the next section including the **Figures 4-6.**

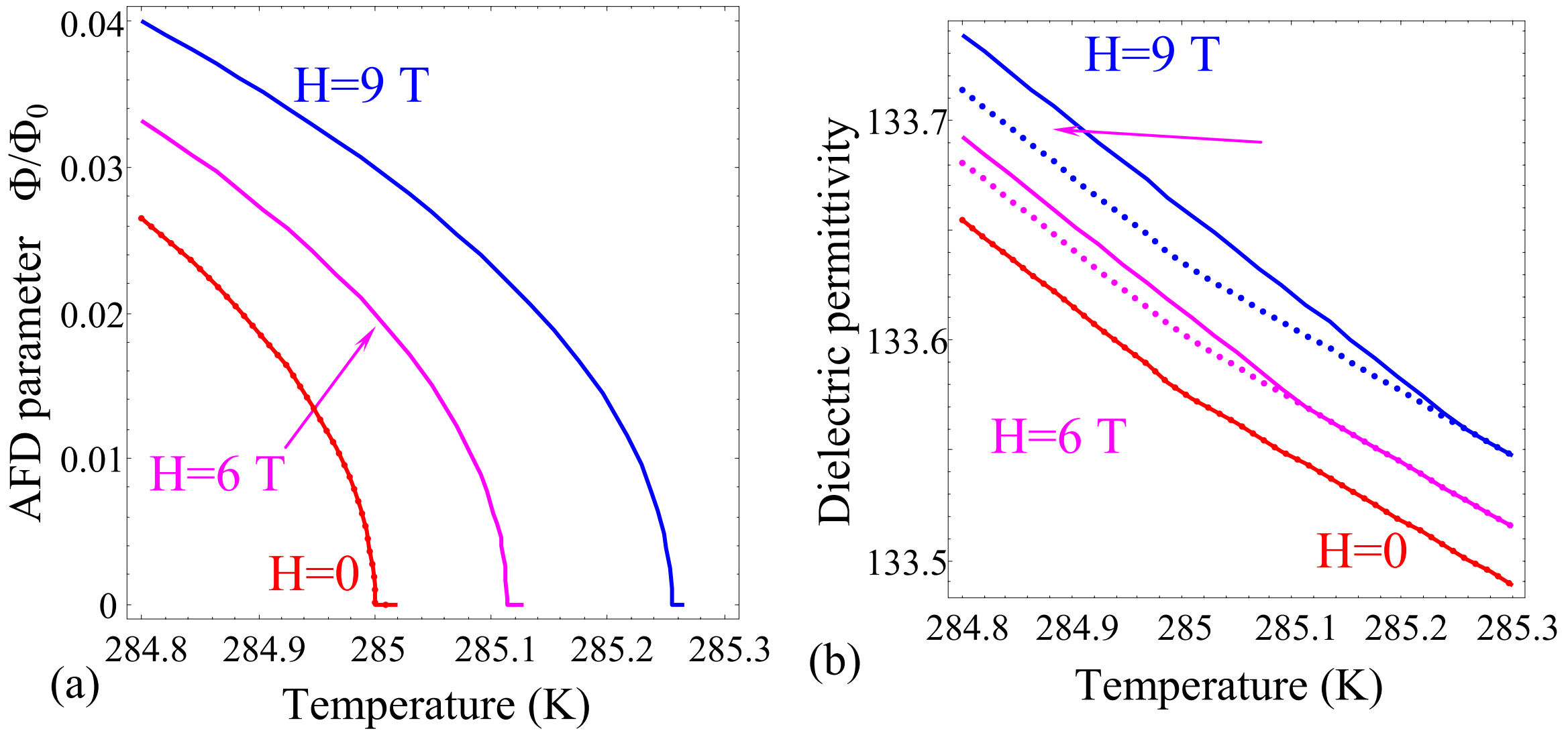

**Figure 3**. Temperature dependence of **(a)** AFD order parameter and **(b)** dielectric permittivity at different external magnetic field values (shown near the curves) near the transition to AFD phase. Solid and dotted curves are calculated with and without RM coupling respectively.

## IV. Strong RM effects at lower temperatures

### IV.1. External field control of the phase diagram

**Figures 4a** and **4b** illustrate the influence of external magnetic field on ferroic material (prototype is ETO) magnetic properties at zero electric field (compare with results of Ref.[23]). **Figures 4c** and **4d** show the influence of external electric field on the ferroic magnetic properties at zero magnetic field. In all panels of the **Figure 4** the cases with and without RM coupling are considered in the temperature region $T<T_S$, so that AFD phase is always present here. Phase diagrams contain antiferrodistortive-paramagnetic (AFD-PM), antiferrodistortive-ferromagnetic (AFD-FM) and antiferrodistortive-antiferromagnetic (AFD-AFM) phases separated by solid curves, which in fact should be interpreted as the dependences of the **critical magnetic** (for the panels **4a** and **4b**) and **electric** (for the panels **4c** and **4d**) fields dependences on temperatures. One can see the strong difference between the phase diagrams calculated with ($\xi_{RM}^{M}=-\xi_{RM}^{L}\neq 0$) and without ($\xi_{RM}^{M}=\xi_{RM}^{L}=0$) RM-coupling. The difference is mainly in the numerical values of the critical fields and temperatures, since the RM-coupling strongly renormalizes magnetic Neel temperature (at about two tens of Kelvin degrees).

At zero electric field RM coupling shifts AFM phase boundary to the lower temperatures region from 26 K to 5.5 K as follows from the **Figures 4a** and **4b**. Namely, the seeding Neel temperature calculated without RM coupling (i.e. without inclusion of the tilt influence on magnetization) is about 26 K for ETO. With RM coupling the Neel temperature appeared equal to experimental value 5.5 K. The RM constant ($2.3\times10^{-16}$ N/(m$^2$A$^2$)) was obtained from the extrapolation of the experimental ferromagnetic susceptibility temperature dependencies.

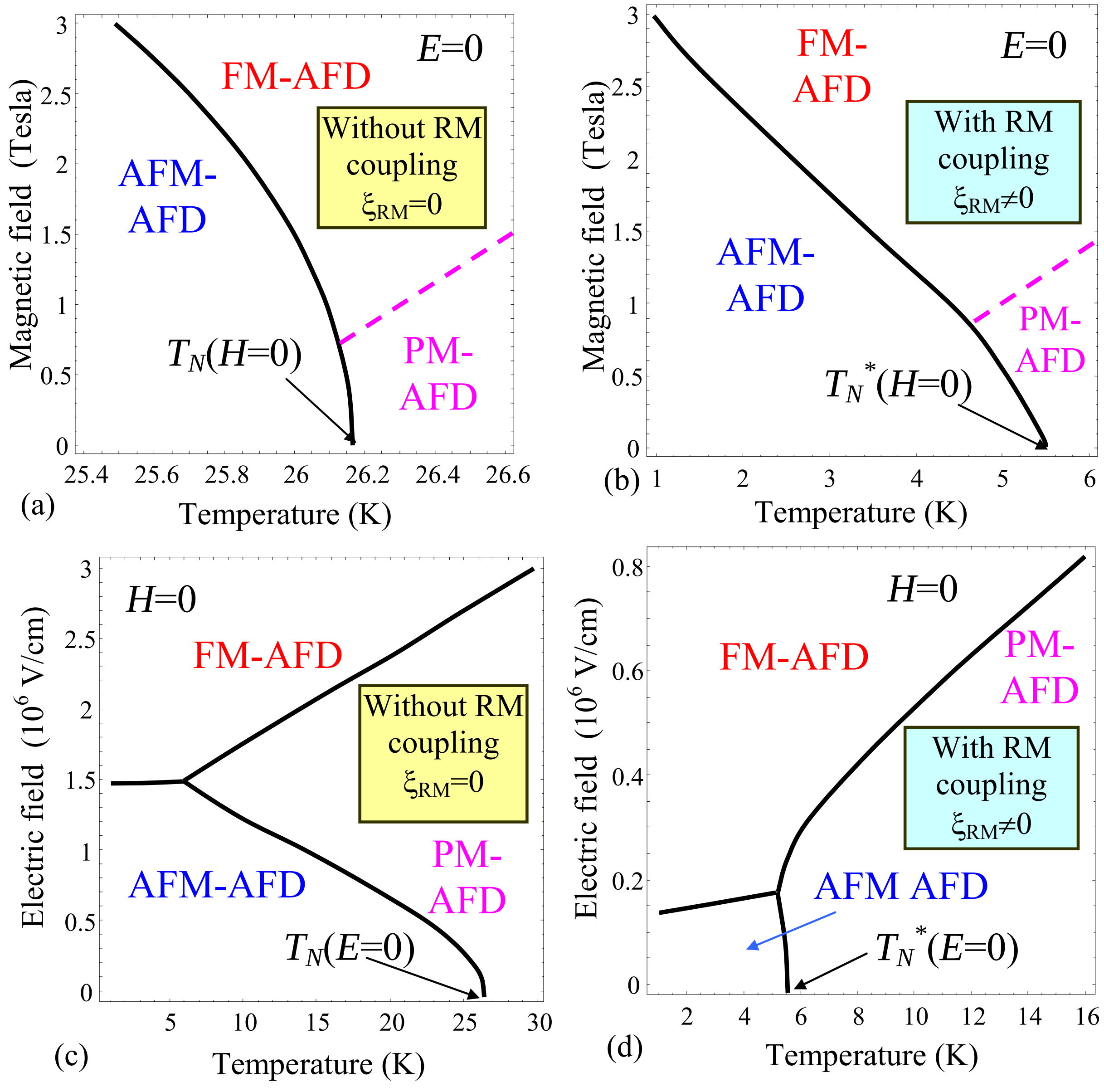

**Figure 4**. **External field control of multiferroic properties.** Phase diagram in the coordinates of temperature versus external magnetic **(a,b)** and electric **(c,d)** fields. Plots **(a)** and **(c)** are calculated without RM coupling ($\xi^M_{RM} = \xi^L_{RM} = 0$). Plots **(c)** and **(d)** are calculated with RM-coupling ($\xi^M_{RM} = -\xi^L_{RM}$, $\xi^L_{RM} = 2.3\times10^{-16}$ N/(m$^2$A$^2$)). Abbreviations PM, FM and AFM denote paramagnetic, ferromagnetic and antiferromagnetic phases respectively. AFD stands for the antiferrodistortive ordered phase.

At zero magnetic field phase diagrams contain three-critical point between FM, AFM and PM phases (see **Figure 4c** and **4d)**. The boundary between FM and AFM phases is an almost straight line. RM coupling changes the phase boundaries and shifts the three-critical point to the lower temperatures and essentially decreases (in about 8 times) the value of the critical electric field.

All the panels of **Figure 4** illustrate the possibility to govern the phase diagram with AFM, FM and PM phases inside AFD phase by external magnetic and electric fields, the latter phenomena originating from RM, RE and ME couplings.

### IV.2. Roto-effect influence on the properties

The temperature dependence of the AFM, AFD order parameters, magnetic and dielectric permittivity at zero external fields ($E$=$H$=0) are demonstrated in the **Figures 5** for different values of RM coupling constant.

**Figure 5a** illustrates the temperature dependence of the AFM order parameter at different values of RM coupling constant $\xi^L_{RM}$. Its increase strongly decreases the critical (Neel) temperature of the AFM order parameter appearance. The transition to AFM phase takes place in the temperature range 6 – 26 K depending on the RM coupling strength. The PM-AFM phase transition is of the second order for zero and small values of $\xi^L_{RM}$ (curves 1-3 in **Figure 4a**). At the sufficiently high values of $\xi^L_{RM}$ the phase transition becomes of the first order (curve 4 in **Figure 5a**).

As one can see from the **Figure 5b,** the appearance of AFM ordering has no effect on the AFD ordering in the absence of RM coupling, since the curve 1 remains straight in comparison with other curves 2-4, which have pronounced additional feature (sharp cusp followed by smooth minima and gradual increase further) at the PM-AFM transition point. In fact, curves 2-4 in the **Figure 5b** illustrate the appearance of the jump-like cusp of the tilt Φ. The nature of the cusp is the sharp increase of the AFD parameter due to the RM-coupling with the sharp changes of $l^2$ (shown in the **Figure 5a**). The increase of RM coupling leads to the cusp height growth as the additional feature on the temperature dependence of AFD order parameter (see curves 2-4 in the **Figure 5b**).

The changes of the magnetic permittivity temperature dependence with RM coupling increase are shown in the **Figure 5c**. Without RM-coupling ( $\xi^L_{RM}=0$ ) the magnetic susceptibility has a very small cusp at the Neel temperature 26 K (curve 1). The cusp increases essentially and shifts to the lower temperatures (up to 5 K) with $\xi^L_{RM}$ increase (curves 2-4). It should be underlined the giant increase (in 100 times) of the permittivity maximum with $\xi^L_{RM}$ increase from 0 to $2.3\times10^{-16}$ SI units.

Typical changes of the dielectric permittivity temperature dependence with RM coupling increase can be seen from the **Figure 5d**. The dielectric permittivity has the jump-like cusp at Neel temperature. RM coupling shifts the position of the cusp from 26 K at $\xi_{RM}^{L}=0$ (curve 1) to 5 K at $\xi_{RM}^{L}=2.3\times10^{-16}$ N/(m$^2$A$^2$) (curve 4). The permittivity moderately increases with $\xi_{RM}^{L}$ increase, but the cusp height increases more significantly.

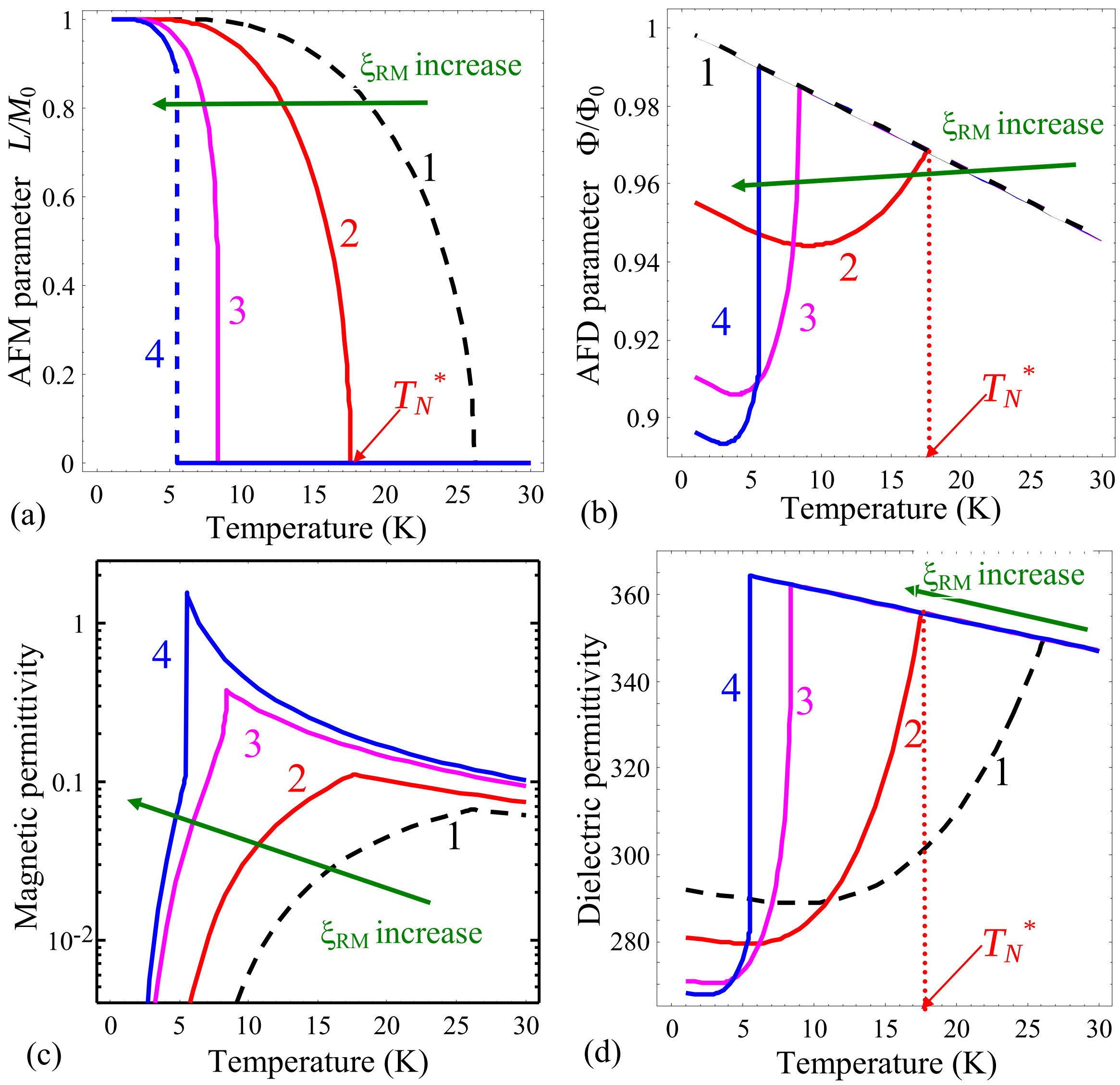

**Figure 5**. **RM coupling influence on the order parameters and permittivities.** Temperature dependence of **(a)** AFM and **(b)** AFD order parameters, **(c)** magnetic and **(d)** dielectric permittivity calculated at low temperatures, zero external fields, $E=H=0$ for $\xi_{RM}^{M}=-\xi_{RM}^{L}$ and different values of RM-coupling coefficient $\xi_{RM}^{L}$ =0, 1, 2 and $2.3\times10^{-16}$ N/(m$^2$A$^2$) (curves 1-4). Other parameters correspond to prototype ETO.

Let us proceed with consideration of external magnetic field influence on the properties with and without RM coupling contribution. Magnetic field causes the changes of magnetization via the field-induced AFM-FM phase transition (see **Figure 4**), while only AFM order exists at $H$=0. With $H$ increase $l$ sharply disappears. **Figure 6** shows how the temperature dependence of AFM order parameter and dielectric susceptibility changes in the magnetic field. Two cases are shown, namely in the absence of RM coupling ($\xi_{RM}^{L}=0$, plots **a** and **c**); and $\xi_{RM}^{L}=2.3\times10^{-16}$ N/(m$^2$A$^2$) **(b, d)**.

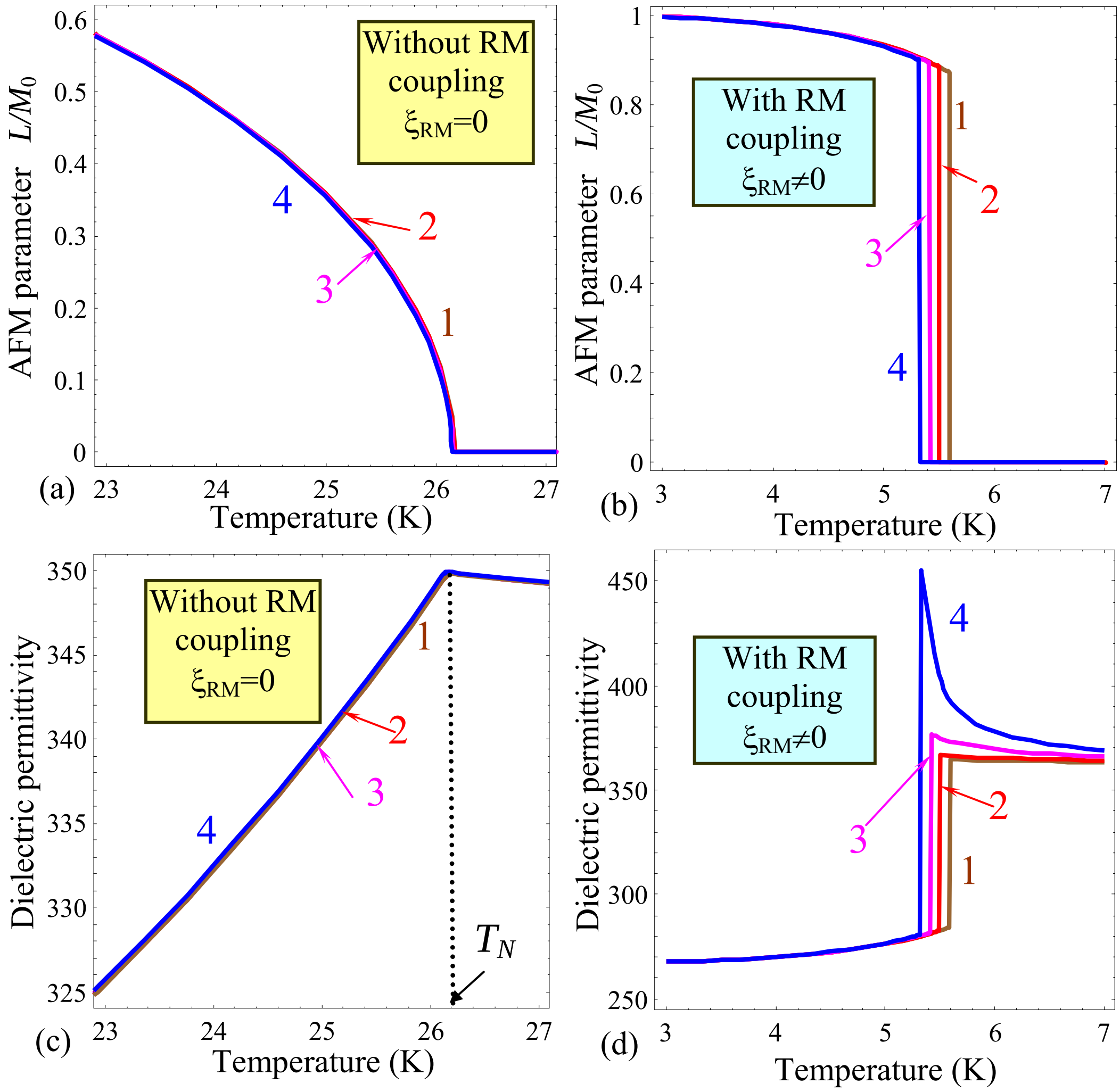

**Figure 6**. **RM coupling changes the transition order.** Temperature dependence of **(a, b)** AFM order parameter and **(c, d)** dielectric permittivity at low temperatures near the transition to AFM phase calculated at zero electric field and different values of RM coupling coefficient $\xi_{RM}^{M}=-\xi_{RM}^{L}=0$ **(a, c)** and $\xi_{RM}^{L}=2.3\times10^{-16}$ N/(m$^2$A$^2$) **(b, d)**. Other parameters quantitatively describe ETO. Curves 1-4 correspond to different values of magnetic field, namely 0, 0.1, 0.2 0.3 Tesla **(b,d)** and 0, 0.2, 0.3 0.6 Tesla **(a,c)**.

AFM order parameter demonstrates the second order phase transition at 26 K without RM-coupling ($\xi_{RM}^{L}=0$), and it is independent on the magnetic field (curves 1-4 almost coincide in the **Figure 6a**). AFM order parameter demonstrates the first order phase transition at 5.5 K and $\xi_{RM}^{L}$ = 2.3×10$^{-16}$ N/(m$^2$A$^2$), at that it becomes dependent on the magnetic field in the transition region (see curves 1-4 in the **Figure 6b**).

Dielectric permittivity is almost independent on magnetic field without RM coupling (see curves 1-4 in the **Figure 6c**). The fact proves a weak contribution of ME coupling in this temperature interval. The cusp at 26 K corresponds to the AFM order appearance. With RM coupling dielectric permittivity has a sharp maximum at 5.5 K and becomes magnetic field dependent in the vicinity of the transition region (see curves 1-4 in the **Figure 6d**).

All results depicted in the **Figures 5** and **6** were obtained on the basis of quantitative calculations with the help of Eqs.(1c)-(1e). To make more clear the physical reasons of the changes of AFM-PM phase transition order, the shift of Neel temperature and other results it appeared useful to look for analytical formulas by minimization of Eq.(1a) with respect to Eqs.(1b)-(1e). Note, that in the case of Eq.(1d) we had to expand entropy into the series over order parameters, so that we obtained both exact and approximate expressions. The details of calculations are listed in the **Supplement**. In particular we obtained that the coefficient before $l^4$ changes its sign under the condition $\frac{M_0^2}{\beta_\Phi}\left(\xi_{RM}^{L}\right)^2 > \frac{\alpha_M^{(T)}}{3}T$. Estimations on the help of the **Table 1** had shown the validity of the inequality. Renormalized Neel and Curie temperatures can be obtained in the form:

$$T_N^* = \frac{T_N - \alpha_M^{(T)}\xi_{RM}^{L}\left(\alpha_T^{(\Phi)}/\beta_\Phi\right)T_S}{1-\alpha_M^{(T)}\xi_{RM}^{L}\left(\alpha_T^{(\Phi)}/\beta_\Phi\right)}, \qquad T_C^* = \frac{T_C - \alpha_M^{(T)}\xi_{RM}^{M}\left(\alpha_T^{(\Phi)}/\beta_\Phi\right)T_S}{1-\alpha_M^{(T)}\xi_{RM}^{M}\left(\alpha_T^{(\Phi)}/\beta_\Phi\right)} \quad (2)$$

Here we used linear temperature dependence of the coefficient $\alpha_\Phi$. Dependences of $T_N^*$ and $T_C^*$ on RM coupling coefficient $\xi_{RM}^{L}$ are shown in the **Figure 7**.

Since the Eq.(2) describes exactly $T_N^*$ and $T_C^*$ it is not the surprise, that the values of $T_N^*$ in **Figures 5** for several RM coupling values perfectly fit the straight line for $T_N^*$ in the **Figure 7.**

In the **Figure 7** special attention has to be paid to the vertical line crossing the point $\xi_{RM}^{L}$ = 2.45×10$^{-16}$ N/(m$^2$A$^2$), where $T_N^* = T_C^*$. As the matter of fact this line is the boundary between AFM and FM phases. The latter is absent in ETO, because its RM coefficient, $\xi_{RM}^{L}$ = 2.3×10$^{-16}$ N/(m$^2$A$^2$), is smaller than the value of the crossing point. Allowing for that $\xi_{RM}^{L}$ value is a characteristic feature of a material, one has to look for the materials with $\xi_{RM}^{L}$ > 2.5×10$^{-16}$ N/(m$^2$A$^2$), that belongs to the group of paraelectric antiferromagnets with AFD phase. Generally speaking this will give the way to obtain the new group of ferromagnetic materials with unusual magnetic properties, which can be interesting for fundamental studies and useful for new applications.

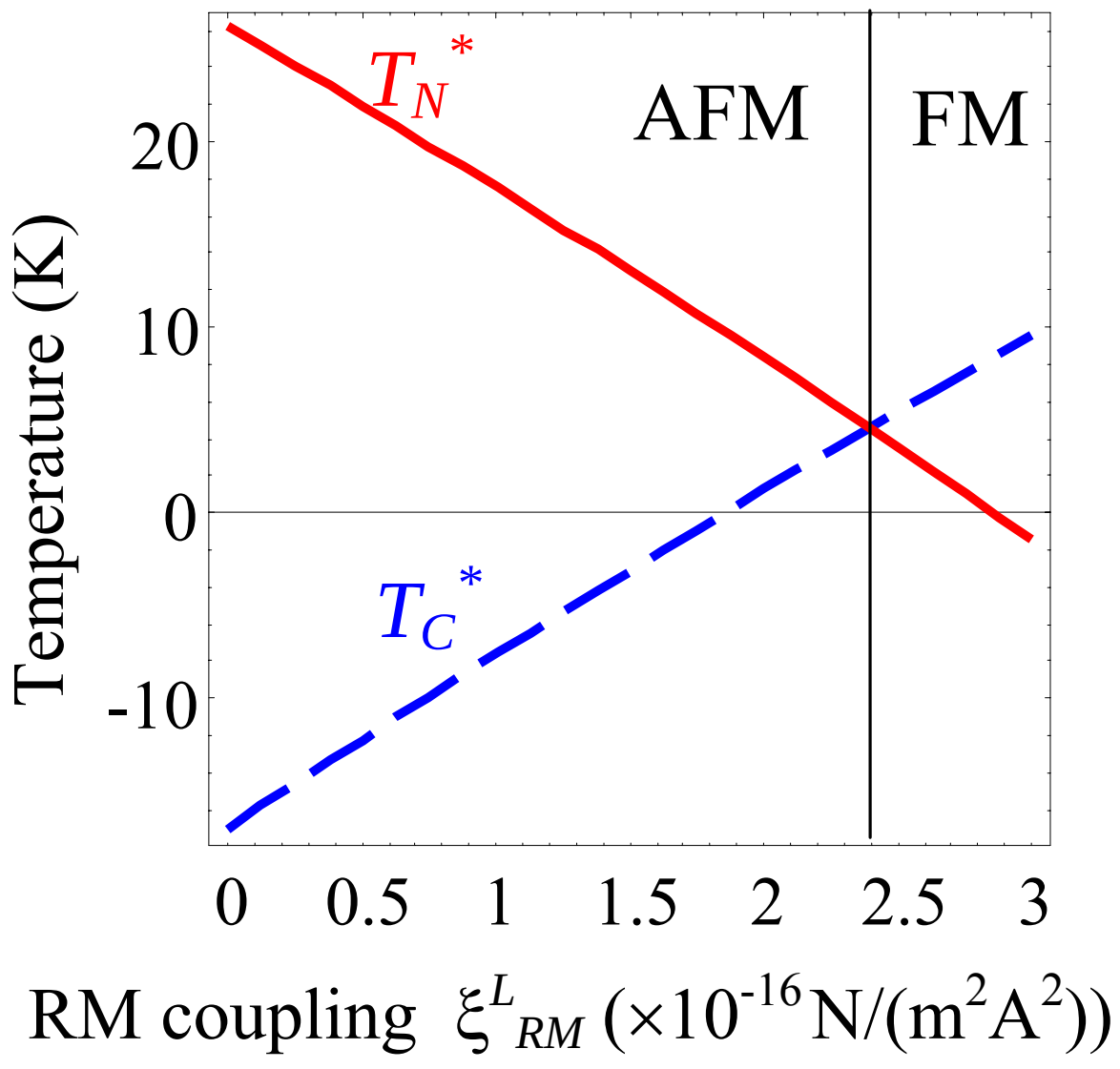

**Figure 7.** Dependences of $T_N^*$ and $T_C^*$ on RM coupling coefficient $\xi^L_{RM}$ calculated at $\xi^M_{RM} = -\xi^L_{RM}$ .

## V. Discussion

Although we performed consideration for incipient ferroelectric ETO with known parameters and coefficients of the LGD-potential expansion, the obtained results can be applied to any paraelectric antiferromagnet of perovskite structure with AFD transition. The statement follows from the approach generality based on the considered form of the free energy (1). For the group of such materials (that can include e.g. $Sr_xEu_{1-x}TiO_3$, $x < 0.2$ [16], where RM coupling value can be different from that in $EuTiO_3$) the ferromagnetic phase can be induced by external electric field that shifts ions and so changes magnetic exchange interaction leading to the transformation of AFM to FM phase at electric field exceeding critical value [23]. Existence of RM coupling opens additional mechanism of AFM to FM phase transformation that decreases essentially (about 7-8 times) the critical field value (compare **Figures 2c** and **2d**). The fact can be important since it opens the opportunities for the control of magnetic phases by small enough electric field.

The main attention in the paper was paid to the influence of RM effect on the properties, such as AFD and AFM transition temperatures, order parameters, magnetic and dielectric permittivity in a broad temperature region. The comparison of these properties with and without RM coupling has clearly shown the existence of the temperature region (T<30 K), where RM effect defines the main features of these properties. Especially important is the changing of Neel temperature that could be higher (about 26 K) without RM coupling than the observed one (5.5

K) with RM coupling; besides the second order transition without RM coupling transforms into the first order one with RM coupling, as it is clearly seen from the **Figures 4** and **5**. Also it is very interesting that RM coupling influences strongly the shape and intensity of magnetic and dielectric permittivity in dependence on the coupling strength and magnetic field value. The field influence on the dielectric permittivity originates from both RM coupling (proportional to $\Phi^2M^2$) and RE coupling (proportional to $\Phi^2P^2$), because changing of $\Phi$ under magnetic field leads to the polarization and so dielectric permittivity changes. The influence of ME coupling on dielectric permittivity proportional to $M^2P^2$ for high magnetic field (H>1 T in our case) is essential also.

Without RM coupling the type of AFM phase transition, the value of Neel temperature, as well as magnetic permittivity anomalous shape are defined by Eqs.(1c)-(1d), which describe complex interplay between magnetic moments ordering including possible nonlinear effects at different temperatures. In the paramagnetic phase the RM coupling influences only slightly the AFD transition temperature and dielectric susceptibility, because of the smallness of magnetic field induced magnetization. Here we report only about the small shift of AFD transition temperature in magnetic field. Essential influence at lower temperatures on the properties as well as on the shape and value of magnetic permittivity speaks in favor of the statement that the type and degree of magnetic ions ordering also depends on RM coupling. Therefore the physical mechanisms included into Eqs.(1c)-(1e) are responsible for the features of the properties predicted in the paper. Because all of them depend on RM coupling strength the observation of predicted phenomena in ETO and other paraelectric antiferromagnets with AFD phase would be extremely desirable. Note that it is not excluded that RE coupling in the incipient ferroelectric STO can be the main reason of ferroelectric phase absence in the material. The statement is in agreement with earlier works of Tagantsev et al [4].

## VI. Conclusions

The influence of RM coupling on the properties of incipient ferroelectrics in temperature regions with coexistence of AFD long-range order and PM phase (the first region, $T_N \leq T \leq T_S$) and AFD long-range order and magnetic order (the second region, $0 \leq T \leq T_N$) was investigated in the LGD theory framework. The main driving force of RM effects was shown to be biquadratic coupling between AFD and AFM order parameter in the second region or with AFD order parameter and magnetization induced by magnetic field in the first region. In the first region RM coupling describes pretty good the influence of magnetic field on AFD transition temperature and magnetic susceptibility observed recently in ETO incipient ferroelectric. However, quantitatively this influence appeared to be very small (few degrees of Kelvin for the transition temperature shift).

RM effect influence on the properties appeared much stronger in the second region with coexistence of two abovementioned long-range orders, i.e. in the multiferroic state. In particular the value of observed Neel temperature was shown to be defined by RM coupling. Without the coupling Neel temperature $T_N$=26 K, and it decreases linearly to observed in ETO value 5.5 K with the coupling increase, while magnetic Curie temperature increases also linearly from $T_C = -17$ K to the value larger than 5.5. K at RM coupling coefficient $\xi^{L}_{RM} > 2.45\times10^{-16}$ N/(m$^2$A$^2$). Therefore the possibility of transformation from AFM to FM phase transition appears in the material with RM coupling larger than that in ETO.

The RM coupling opens the way to control the form of phase diagrams by application of external magnetic and electric fields. It is worth to underline that the value of critical electric field required to induce the transition from AFM phase to FM phase appeared to be essentially smaller than the one calculated without RM coupling. The latter fact seems to be important for applications. Note the appearance of the three-critical point in antiferrodistortive phase region under electric field application, namely FM, AFM and PM phases coexist in the point.

Without external fields the properties depend essentially on RM coupling values. In particular for small or high enough RM coupling values the phase transition becomes of the second or the first order respectively. Among other interesting anomalies we would like to underline the anomalies of the temperature dependence of magnetic and dielectric permittivities with RM coefficient increase. Namely we forecasted the giant increase (in 100 times) of magnetic permittivity maximum height with RM coupling increase from 0 to $2.3\times10^{-16}$ N/(m$^2$A$^2$). Obtained cusp-like shape of dielectric permittivity at Neel temperature shifts moderately at lower temperatures at RM coupling constant increase, but the height of the cusp increases more significantly. It is worth to underline that AFM order parameter and dielectric permittivity of ETO are independent of magnetic field value at low temperatures for zero RM coupling constant, and the dependence appears only at nonzero RM constant. All the predicted new effects originated from the strong influence of RM coupling on the properties and phase diagrams are waiting for experimental verification.

## Supplement

The transition order changes because biquadratic RM coupling term in Eq.(1e), namely:

$$\Delta g_{Coupling}=\frac{M_0^2}{2}\Phi^2\left(\xi_{RM}^M m^2+\xi_{RM}^L l^2\right), \qquad (S.1)$$

changes the nonlinearity structure of the system AFM energy, namely:

$$\begin{aligned}&\frac{\alpha_M^{(T)}M_0^2}{2}\left[(T-T_C)m^2+(T-T_N)l^2\right]+\frac{\alpha_M^{(T)}M_0^2}{2}T\left(S(m+l)+S(m-l)-m^2-l^2\right)\\&\approx\frac{\alpha_M^{(T)}M_0^2}{2}\left[(T-T_C)m^2+(T-T_N)l^2\right]+\frac{\alpha_M^{(T)}M_0^2}{2}T\left(m^2l^2+\frac{m^4+l^4}{6}+...\right)\end{aligned}. \qquad (S.2)$$

Without RM coupling corresponding nonlinear terms in Eq.(S.1a) are always positive in the ordered phase, i.e. the phase transition must be the second order one. Allowing for RM effect contribution in Eq.(S.1), Eq.(1c) at $P$=0 can be rewritten as

$$g_{AFD}^*=\frac{\alpha_\Phi^*}{2}\Phi^2+\frac{\beta_\Phi}{4}\Phi^4, \qquad (S.3)$$

where $\alpha_\Phi^*=\alpha_\Phi(T)+M_0^2\left(\xi_{RM}^M m^2+\xi_{RM}^L l^2\right)$ and so that $\Phi^2=-\alpha_\Phi^*/\beta_\Phi$ . So that is gives equilibrium tilt value $\Phi^2=\Phi_{\xi 0}^2-M_0^2\left(\xi_{RM}^M m^2+\xi_{RM}^L l^2\right)/\beta_\Phi$ , where $\Phi_{\xi 0}^2=-\alpha_\Phi/\beta_\Phi$ is the tilt value in the absence of rotomagnetic coupling.

Now we can summate all the terms proportional to $l$ and $m$ powers in Eqs.(S.1)-(S.3): and obtain

$$\Delta\left(TS+g_{AFD}^*\right)\equiv TS+\frac{\alpha_\Phi^*}{2}\Phi^2+\frac{\beta_\Phi}{4}\Phi^4=\left(\begin{array}{l}\frac{\alpha_M^{(T)}M_0^2}{2}\left[(T-T_C)m^2+(T-T_N)l^2\right]\\+\frac{\alpha_M^{(T)}M_0^2}{2}T\left(m^2l^2+\frac{m^4+l^4}{6}\right)+\\\frac{\alpha_\Phi^*}{2}\left(\Phi_{\xi 0}^2-\frac{M_0^2}{\beta_\Phi}\left(\xi_{RM}^M m^2+\xi_{RM}^L l^2\right)\right)\\+\frac{\beta_\Phi}{4}\left(\Phi_{\xi 0}^2-\frac{M_0^2}{\beta_\Phi}\left(\xi_{RM}^M m^2+\xi_{RM}^L l^2\right)\right)^2\end{array}\right). \qquad (S.4)$$

After cumbersome but elementary transformations Eq.(S.4) can be identically rewritten as

$$\begin{aligned}&\frac{\alpha_M^{(T)}M_0^2}{2}\left[(T-T_C)m^2+(T-T_N)l^2\right]+\frac{M_0^2\Phi_{\xi 0}^2}{2}\left(\xi_{RM}^M m^2+\xi_{RM}^L l^2\right)+\\&\left(\alpha_M^{(T)}M_0^2T-\frac{M_0^4}{\beta_\Phi}\xi_{RM}^M\xi_{RM}^L\right)\frac{m^2l^2}{2}+\left(\frac{\alpha_M^{(T)}M_0^2}{3}T-\frac{M_0^4}{\beta_\Phi}\left(\xi_{RM}^M\right)^2\right)\frac{m^4}{4}+\left(\frac{\alpha_M^{(T)}M_0^2}{3}T-\frac{M_0^4}{\beta_\Phi}\left(\xi_{RM}^L\right)^2\right)\frac{l^4}{4}\end{aligned}$$

(S.5)

Finally one can obtain the coefficient before $l^4$ in the form

$$\frac{M_0^2}{4}\left(\frac{\alpha_M^{(T)}}{3}T - \frac{M_0^2}{\beta_\Phi}\left(\xi_{RM}^L\right)^2\right) \tag{S.6}$$

Since $\beta_\Phi > 0$ the coefficient before $l^4$ changes its sign under the condition $\frac{M_0^2}{\beta_\Phi}\left(\xi_{RM}^L\right)^2 > \frac{\alpha_M^{(T)}}{3}T$. Estimations on the help of the **Table 1** had shown the validity of the inequality. Moreover, since we regard that $\xi_{RM}^L = -\xi_{RM}^M$, the coefficient before $m^4$ changes its sign under the same condition in accordance with Eq.(S.5). Such behavior is a typical parabolic instability in a system with several order parameters.

In order to derive renormalization of Neel and Curie temperatures we gather the first terms proportional to $l^2$ and $m^2$ in Eq.(S.5), namely $\frac{\alpha_M^{(T)}M_0^2}{2}\left[\left(T - T_C + \frac{\xi_{RM}^M}{\alpha_M^{(T)}}\Phi_{\xi 0}^2\right)m^2 + \left(T - T_N + \frac{\xi_{RM}^L}{\alpha_M^{(T)}}\Phi_{\xi 0}^2\right)l^2\right]$ and obtain that:

$$T_N^* = T_N - \frac{\xi_{RM}^L}{\alpha_M^{(T)}}\Phi_{\xi 0}^2\left(T_N^*\right) \tag{S.7a}$$

$$T_C^* = T_C - \frac{\xi_{RM}^M}{\alpha_M^{(T)}}\Phi_{\xi 0}^2\left(T_N^*\right) \tag{S.7b}$$

Where $\Phi_{\xi 0}^2\left(T_N^*\right) = -\alpha_\Phi\left(T_N^*\right)/\beta_\Phi$. Since $\xi_{RM}^L = -\xi_{RM}^M > 0$, we get $T_N^* < T_N$ and $T_C^* > T_C$.

[11] A. Bussmann-Holder, J. Kohler, R. K. Kremer, and J. M. Law, Phys. Rev. B **83**, 212102 (2011).

[12] M. Allieta, M. Scavini, L. J. Spalek, V. Scagnoli, H. C. Walker, C. Panagopoulos, S. S. Saxena, T. Katsufuji, and C. Mazzoli, Phys. Rev. B **85**, 184107 (2012).

[13] K. Z. Rushchanskii, N. A. Spaldin, andM. Lezaic, Phys. Rev. B **85**, 104109 (2012).

[14] V. Goian, S. Kamba, O. Pacherova, J. Drahokoupil, L. Palatinus, M. Dusek, J. Rohlıcek, M. Savinov, F. Laufek, W. Schranz, A. Fuith, M. Kachlık, K. Maca, A. Shkabko, L. Sagarna, A.Weidenkaff, and A. A. Belik, Phys. Rev. B **86**, 054112 (2012).

[15] A. P. Petrovic, Y. Kato, S. S. Sunku, T. Ito, P. Sengupta, L. Spalek, M. Shimuta, T. Katsufuji, C. D. Batista, S. S. Saxena, and C. Panagopoulos, Phys. Rev. B **87**, 064103 (2013).

[16] E. A. Eliseev, M. D. Glinchuk, V. V. Khist, C.-W. Lee, C. S. Deo, R. K. Behera, and A. N. Morozovska, J. Appl. Phys. **113**, 024107 (2013).

[17] Anna N. Morozovska, Maya D. Glinchuk, Rakesh K. Behera, Basyl Y. Zaylichniy, Chaitanya S. Deo, Eugene A. Eliseev. Phys.Rev. **B 84**, 205403 (2011)

[18] A. N. Morozovska, Y. Gu, V. V. Khist, M. D. Glinchuk, L.-Q. Chen, V. Gopalan, and E. A. Eliseev, Phys. Rev. B **87**, 134102 (2013).

[19] Z. Guguchia, A. Shangelaya, H. Keller, J. Kohler, and A. Bussmann-Holder, Phys. Rev. B **85**, 134113 (2012).

[20] S.V. Vonsovskii, Magnetism (John Wiley and Sons, New York, 1974).

[21] J. Lee, L. Fang, E. Vlahos, X. Ke, Y. Jung, L. Kourkoutis, J. Kim, P. Ryan, T. Heeg, M. Roeckerath, V. Goian, M. Bernhagen, R. Uecker, P. Hammel, K. Rabe, S. Kamba, J. Schubert, J. Freeland, D. Muller, C. Fennie, P. Schiffer, V. Gopalan, E. Johnston-Halperin, and D. Schlom, Nature (London) **466**, 954 (2010).

[22] It should be noted, that while considering experimental results on magnetic susceptibility, we recalculated it from "cm$^3$/mol" units to dimensionless one using the following factor $1/\left(N_A a_0^3\right)$, where $N_A$=6.022*10$^{23}$ 1/mol is the Avogadro number and $a_0 = 3.905$ Å is the lattice constant of ETO at room temperature.

[23] Maya D. Glinchuk, Eugene A. Eliseev, Yijia Gu, Long-Qing Chen, Venkatraman Gopalan and Anna N. Morozovska. Phys. Rev. **B 89**, 014112 (2014).